# What fuel for a rocket ?

**E. N. Miranda**[*]

CRICYT - CONICET

5500 - Mendoza, Argentina

and

Dpto. de Física

Univ. Nac. de San Luis

5700 - San Luis, Argentina

**Resumen:** Se han utilizado conceptos sencillos de física general y de termodinámica para analizar el problema de la propulsión de un cohete. Haciendo algunas suposiciones razonables se ha encontrado una expresión para la velocidad de expulsión de los gases que muestra cuales son las características deseables de un buen combustible para cohetes.

**Abstract:** Elementary concepts from general physics and thermodynamics have been used to analyze rocket propulsion. Making some reasonable assumptions, an expression for the exit velocity of the gases is found. From that expression one can conclude what are the desired properties for a rocket fuel.

[*] E-mail: emiranda@lab.cricyt.edu.ar

Rockets are usual problems for a student that takes a first course in physics [1, 2] because they are common examples of variable mass systems. However, when the student takes a thermodynamics course, he does not deal with the problem again. This is a regrettable omission because he has the tools to deepen the analysis and draws some conclusions about the desirable properties of a rocket fuel. The aim of this paper is to fill that gap. An elementary knowledge of mechanics and thermodynamics is assumed for the analysis; however the conclusions are interesting and the pedagogical advantage of this issue is clear: it captures the student's attention.

A rocket is one of the simplest engines for imparting motion to a vehicle. Conventional engines first convert chemical energy to heat that is used to drive a steam engine, turbine or internal combustion engine; these in turn drive the vehicle. In a rocket the conversion is more direct: chemical energy heats matter in a rigid chamber to a high temperature; the matter is then ejected through a nozzle in a specified direction. The reaction from the jets pushes the rocket forward. The rocket motor consists of a combustion chamber and a nozzle, and it converts the random thermal motion of the molecules to collimated jets with a defined macroscopic velocity.

Conventional engines are designed to propel a vehicle with constant mass at a constant velocity. On the contrary, the mass of a rocket decreases steadily as its matter is ejected. The rocket should accelerate rapidly during the burning time because once the fuel has been burnt out, the rocket continues in free flight. From this fact, one concludes that the impulse –i.e. the product of the accelerating force times the time during which is exerted- is a relevant quantity to take into account. But a propellant that gives a large impulse only when a large mass of it is burnt is not as useful as one that gives the same impulse burning a smaller mass. For this reason one should consider the specific impulse, which is defined as the impulse per unit mass of propellant. The specific impulse of a fuel is the magnitude that helps the rocket designer to choose the right propellant.

In the ideal case, the expelled gas flow is perpendicular to the nozzle, and the gas exit pressure is the atmospheric one. Therefore, the total force acting $F$ on the vehicle can be evaluated analyzing the change in the momentum due to the gas flow leaving the nozzle.

$$F = \frac{d(mv)}{dt} = v\dot{m} \tag{1}$$

where $\dot{m}$ is the mass of the expelled gas per unit time and $v$ is its velocity. If one assumes that the engine has been on for a time $t_b$ and a mass of propellant $M_p$ is used up, then the specific impulse $I$ is:

$$I = \frac{F\, t_b}{M_p} = \frac{v\, \dot{m}\, t_b}{\dot{m}\, t_b} = v \tag{2}$$

This means that the specific impulse of a propellant is equal to the velocity of the gases leaving the nozzle.

The previous considerations are valid for the ideal case; for a real situation the force acting to propel the rocket is:

$$F = \lambda\, \dot{m}\, v_e + (p_e - p_o)\sigma \tag{3}$$

$\lambda$ is a factor less than unity to correct for non-perpendicular flow at the exit. On the other hand, $(p_e-p_o)\sigma$ represents the "pressure thrust" exerted across the nozzle because of the difference between pressure of the exhaust gases $p_e$ and the atmospheric pressure $p_o$; $\sigma$ is the nozzle cross section. Usually, this second term is much smaller than the first one of expression (3). Additionally, we are interested in the thermodynamics of the rocket fuel and not in questions regarding rocket design. For those reasons, the second term of expression (3) is disregard in what follows.

Let us assume that in the combustion chamber there is a single gas that is heated for some external agent. Of course, it is heated by its own combustion, but this fact will be taken into account later. The gas reaches a high temperature and pressure, and it expands through the nozzle. For the first law, we may write:

$$dq = dU + w = dU + pdV = dH - V_m dp \tag{4}$$

The meaning of the symbols is the usual one found in any standard textbook on thermodynamics [3]. The behavior of one mol is considered and $V_m$ is the molar volume;

consequently the density is $\rho = M / V_m$ where $M$ is the molecular mass. The enthalpy $H$ is defined as $H = U + pV$.

From hydrodynamics [4], one may write the Euler equation. For a stationary flow it states that:

$$-dp = \rho \, v \, dv \tag{5}$$

This equation relates the change in the flow pressure with the change in velocity. Replacing in (1), and remembering the density definition, one gets:

$$dq = dH + V_m \frac{M}{V_m} v \, dv = d\left(H + \tfrac{1}{2} M v^2\right) \tag{6}$$

It is clear that the second term between the brackets is the kinetic energy *KE* of the expelled gas.

Now, we can make a reasonable assumption: the whole process is fast enough as not to exchange heat with the environment, i.e. *dq = 0* and the process is adiabatic. As a consequence, the enthalpy plus the kinetic energy of the gas is a constant. If we use the subscript *c* for the values of those magnitudes in the combustion chamber and the subscript *e* for the exit values, we get:

$$H_c + \tfrac{1}{2} M v_c^2 = H_e + \tfrac{1}{2} M v_e^2 \tag{7}$$

Of course, in the combustion chamber the gases have not a definite macroscopic velocity and $v_c = 0$. The gases exit velocity comes out to be:

$$v_e = \sqrt{\frac{2(H_c - H_e)}{M}} \tag{8}$$

Our next step is to relate the enthalpies that appear in (8) with magnitudes that are easy to measure, temperature being a good option for that. One should remember that the specific heat at constant pressure could be written as:

$$C_p = \left(\frac{\partial H}{\partial T}\right)_p \tag{9}$$

Assuming an ideal behavior of the gas $C_p$ is constant, and:

$$H_c - H_e = \int_{T_e}^{T_c} C_p \, dT = C_p (T_c - T_e) \tag{10}$$

Replacing in (8), we get:

$$v_e = \sqrt{\frac{2 C_p (T_c - T_e)}{M}} \tag{11}$$

Equation (11) is a good expression to end up with; however, since we have assumed that the gas is an ideal one and the expansion is adiabatic, we may go ahead a little further. For an ideal gas, the specific heat can be written as:

$$C_p = \frac{\gamma}{1-\gamma} R \tag{12}$$

$R$ is the gas constant and $\gamma$ is the ratio between the gas specific heats that depends on the molecular structure of the gas. On the other hand, for an adiabatic expansion [3] one can write:

$$\frac{T_e}{T_c} = \left(\frac{p_e}{p_c}\right)^{\frac{\gamma-1}{\gamma}} \tag{13}$$

Replacing eqs.(12) and (13) in (11), a final expression for the velocity of the exhaust gases is obtained:

$$I = v_e = \sqrt{\frac{2\gamma R T_c}{(\gamma-1)M} \left[1 - \left(\frac{p_e}{p_c}\right)^{\frac{\gamma-1}{\gamma}}\right]} \tag{14}$$

Up to now, we have considered a single gas that somehow is heated. Now it is time to analyze the process with greater detail. The propellant is ignited in the presence of an oxidant, which is typically oxygen. Therefore, in the combustion chamber there is actually a mixture of gases: the propellant, the oxidant and the products of the combustion. It is the heat liberated by the combustion that rises the temperature of the gas mixture, and this mixture of gases is expelled through the nozzle. For this reason, in eq. (14) M and γ should be understood as an average molecular mass and specific heat ratio.

The rocket designer wants to maximize the specific impulse $I$ of the fuel, and from eq. (14) we may draw some conclusions:

1) The temperature $T_c$ of the combustion chamber should be as high as possible. Therefore, a propellant with a large combustion heat should be chosen.
2) Since $p_e$ is bounded by the atmospheric pressure, the pressure in the combustion chamber should be as high as possible.
3) The molecular mass $M$ of the mixture should be as low as possible.

The requirements 1) and 2) are constrained by the mechanical strength of the combustion chamber, and as usual in engineering problems, a compromise should be reached between the strength and the weight of the chamber. Condition 3) explains why the preferred mixture for rocket engines is hydrogen and oxygen. There is no lighter propellant than hydrogen, and it liberates an appreciable amount of heat when burning in the presence of oxygen.

The problems for a rocket designer are formidable, however, in this article a simple analysis has lead us to valid conclusions. The tools we have used are known by a physics or engineering student that has had his first contact with mechanics and thermodynamics. This article can be discussed in a thermodynamics course as an application that goes beyond the usual problems with heat engines and the students will be motivated because they are doing "rocket science"

**Acknowledgment:** The author is supported by the National Research Council of Argentina (CONICET).